%% file: ms.tex
\documentclass[iop,revtex4,twocolappendix]{emulateapj}

\usepackage{amssymb}
\usepackage[usenames,dvipsnames,svgnames]{xcolor}
\usepackage{amsmath, mathrsfs}
\usepackage{amsfonts}

\definecolor{darkgreen}{rgb}{0,.5,0}

\newcommand{\add}{} 

\newfont{\nf}{cmfib8 at 10pt}

\newcommand{\Rearth}{$R_{\oplus}$}

\newcommand{\kepler}{{\it Kepler}}

\newcommand{\nTdur}{{\textcolor{black}{1.5}}}
\newcommand{\mad}[1]{\ensuremath{ {\rm median( abs(} #1 ))}}

\newcommand{\drtwofour}{Q1-Q17 DR24}

\newcommand{\nKoisTested}{\textcolor{black}{228}}

\newcommand{\nKoisFail}{\textcolor{black}{20}}
\newcommand{\nFaMissed}{\textcolor{black}{10}}

\shortauthors{Mullally et al.}

\begin{document}

\title{Identifying False Alarms in the \kepler\ Planet Candidate Catalog}
\shorttitle{\kepler\ False Alarms}

\author{F.~Mullally\altaffilmark{1},
Jeffery~L.~Coughlin\altaffilmark{1},
Susan~E.~Thompson\altaffilmark{1},
Jessie~Christiansen\altaffilmark{2},
Christopher~Burke\altaffilmark{1},
Bruce~D.~Clarke\altaffilmark{1},
Michael~R.~Haas\altaffilmark{3}
}

\altaffiltext{1}{SETI/NASA Ames Research Center, Moffett Field, CA 94035, USA}
\altaffiltext{2}{NASA Exoplanet Science Instititute, California Institute of Technology, Pasadena, CA 91125}
\altaffiltext{3}{NASA Ames Research Center, Moffett Field, CA 94035, USA}

\email{fergal.mullally@nasa.gov}


\begin{abstract}
We present a new automated method to identify instrumental features masquerading as small, long period planets in the \kepler\
 planet candidate catalog. These systematics, mistakenly identified as planet transits, can have a strong impact on occurrence rate calculations because they cluster in a region of parameter space where \kepler's sensitivity to planets is poor.
We compare individual transit-like events to a variety of models of real transits and systematic events, and use a Bayesian Information Criterion to evaluate the likelihood that each event is real.
We describe our technique and test its performance on simulated data. Results from this technique are incorporated in the \kepler\ Q1-17 DR24 planet candidate catalog of \citet{Coughlin15}.
\end{abstract}

\setlength{\parskip}{1.0ex plus0.5ex minus0.2ex}


\section{Introduction}

One of the major goals of the \kepler\ mission \citep{Koch10} is to estimate the frequency of Earth-size planets in the habitable zones of solar-type stars. To that end, the spacecraft collected 4 years of near-continuous data on $\sim$150,000 stars, searching for the faint signal of small transiting planets.

{\add
The Q1-Q16 catalog of \kepler\ planet candidates \citep{Mullally15cat} reported 554 new candidate planets. They noted an excess of candidates at periods longer than 100\,days over what would be expected if planets were evenly distributed in orbital period.
They identified two large populations of long-period false alarm Threshold Crossing Events (TCEs, or the statistically significant periodic events in a lightcurve that are vetted to produce the planet candidate catalog). One narrow peak centered around 372\,days (the orbital period of the spacecraft), and a larger, broader peak spanning 200-600\,days (see their \S~3.1). While their vetting process filtered out most of the false alarm events from the narrow peak (leaving no excess of candidates at this orbital period), they concluded the observed excess of long-period planet candidates was more likely due to incomplete filtering of the broader peak than any super abundance of small planets in this period range (see their \S~9.1). }

\citet{Burke15} computed the occurrence rates of planets as a function of radius and period around solar analogs using the catalog of \citet{Mullally15cat}.
They found a sharp rise in the computed frequency of Earth analogs for periods longer than 300\,days.
They traced the excess to 5 planet candidates with radii $<1.2$\,\Rearth\ and periods of 450--550\,days, a region of parameter space where such planets could only be detected around 1\% of their sample (see Figure~14 in Burke et al.).

It is clear that understanding the reliability of the \kepler\ catalog (the fraction of candidates actually due to transiting planets) is an important precondition to measuring terrestrial planet occurrence rates.
Much work has been done identifying astrophysical false positives due to eclipsing binaries and stellar multiplicity scenarios \citep[e.g.][]{Torres11, Morton12, Colon15, Santerne12, Bryson13, Desert15}, but less on instrumental artifacts.
The statistical validation techniques used on \kepler\ planets \citep[e.g.][]{Torres11, Rowe14multis} compare the probability that a transit was due to a planet to various eclipsing binary type scenarios.
However, small, long period \kepler\ candidates have a non-negligible probability of being caused by an instrumental or processing artifact.
\citet{Rowe14multis} explicitly avoided low signal to noise cases for this reason.

In this paper we report on a new method to identify and reject false alarm candidates at long periods. This method, which we dub Marshall\footnote{After a character in the animated series Paw Patrol.}${^{\rm ,}}$\footnote{Never let a three year old name your algorithms.}, fits the individual transit events for candidates with models of transits and commonly found artifacts, and evaluates which one is more probable based on a Bayesian Information Criterion. Based on simulations, we find that known artifact types are rejected $\approx$ 60--70\% of the time, while transits are preserved at the 95\%, level as discussed in \S~\ref{performance}.

Marshall is one component of the \kepler\ Robovetter, an automated process for vetting planet candidates in \kepler\ data, which includes the Flux Robovetter \citep{Coughlin15}, the Centroid Robovetter \citep{Mullally15crbv}, the ephemeris matching of \citet{Coughlin14}, {\add and the machine learning technique of \citep{Thompson15} that identifies short period variable stars mistakenly identified as planets}. The Robovetter replaces the manual vetting approach of previous catalogs with an automatic, rules-based technique. In addition to removing some of the inevitable subjectivity of the manual process, the performance of the Robovetter can be tested against large numbers of simulated transits.

Marshall can be easily applied to any transit search where short duration signals (such as instrumental artifacts) are misidentified as transits. In any matched filter approach, such as Box Least Squares \citep{Kovacs02}, or the TPS algorithm used by the \kepler\ pipeline \citep{Seader15}, non-transit signals will be occasionally mistaken for transits. If the false alarm signals can be modeled, the Marshall approach can be used separate true transits from the false alarms. The Marshall algorithm will be useful analyzing data from the upcoming TESS and PLATO missions. We make a reference implementation in Python available at {\url https://sourceforge.net/projects/marshall}.

\section{Method}
The key insight of our technique is that long period candidates  have few enough events that there is sufficient signal-to-noise in an individual event to discriminate between valid and invalid transit shapes. By looking at individual events we have access to information about the transit shape that can be lost in the folded transit event.

For each \kepler\ Object of Interest (KOI) in the \drtwofour\ catalog \citep{Coughlin15} we extract a snippet of data \nTdur\ times the reported transit duration either side of the midpoint of each individual transit event (i.e. once per orbital period of the candidate planet). We use the transit parameters (orbital period, epoch, transit duration) from \citet{Seader15}. We obtain the publicly available lightcurves from MAST\footnote{\url{http://archive.stsci.edu/kepler}} and use the co-trended lightcurve which corrects for  many instrumental features \citep[available in the {\tt PDCSAP\_FLUX} column of the lightcurve file; see Fraquelli \& Thompson 2014 and][]{Smith12}.
\nocite{Fraquelli14}
We then fit each event with the following set of models:

\begin{enumerate}
\item A parabola.

\item A parabola with a negative-going step-wise discontinuity (i.e. a step down) at the reported time of ingress.

\item A parabola with a positive-going step-wise discontinuity at the reported time of egress.

\item A parabola plus a Sudden Pixel Sensitivity Dropout event (SPSD; an SPSD is typically caused by a cosmic ray hit on the {\sc ccd}).  We model the SPSD shape as

\begin{equation}
\left\{ \begin{array}
{r@{\quad:\quad}l}
0 & t < t_{\rm ingress} \\
-A e^{ -\tau (t-t_{\rm ingress})} &  t > t_{\rm ingress} \\
\end{array}
\right.
\end{equation}

where $A$ and $\tau$ are free parameters, and $t_{\rm ingress}$ is the reported transit ingress time.

\item A parabola plus a box shaped transit, defined as
\begin{equation}
\left\{ \begin{array}
{r@{\quad:\quad}l}
-d &  t_{\rm ingress} < t < t_{\rm egress} \\
0 &  {\rm otherwise} \\
\end{array}
\right.
\end{equation}

where $d$ is a free parameter, while $t_{\rm ingress}$ and $t_{\rm egress}$ are constrained so that the transit mid-point is fixed.

\end{enumerate}

We include a parabolic term in each model to describe the continuum flux (i.e the flux we would expect if the proposed transit were not present). 
The algorithm used for co-trending \citep[PDC,][]{Smith12} tries to preserve stellar variability, so the continuum is often not flat even on the short time scales of interest here.

\begin{figure}
     \begin{center}
    \includegraphics[angle=0, scale=.37]{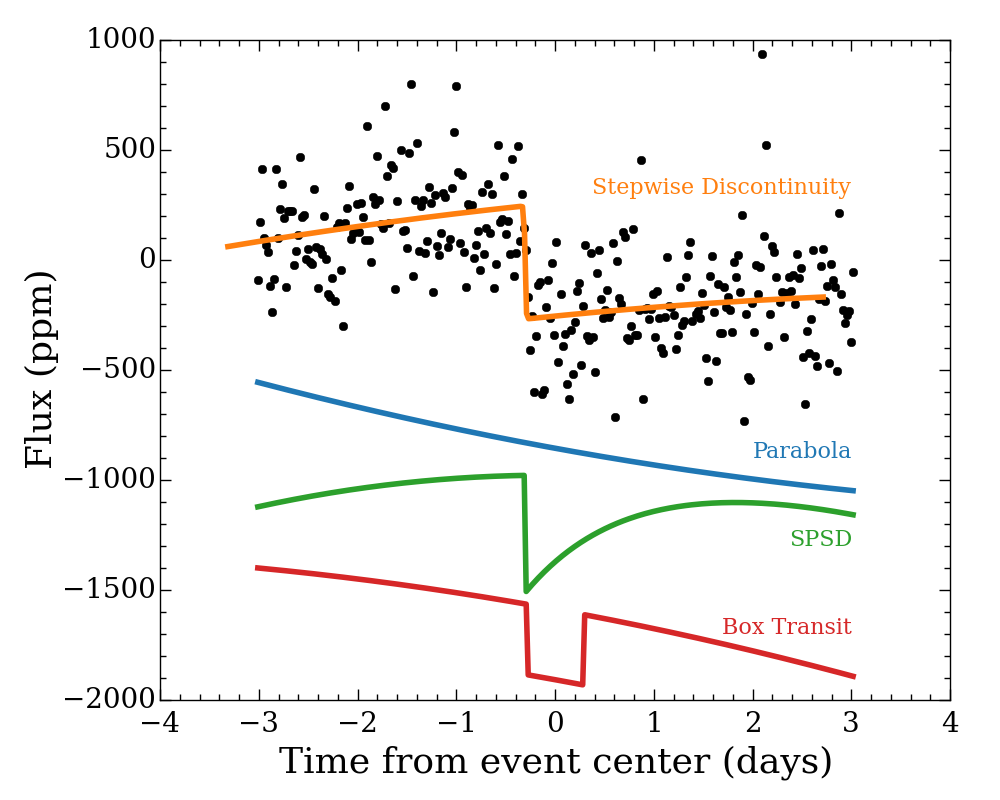}
    \caption{An example instrumental feature from \kepler\ data (Kepler Id 4575824, alias K06428.01), showing the functional forms of the models fit to the event, offset vertically for clarity. The model in each case includes a parabola to describe the shape of the continuum.
\label{example}}
     \end{center}
 \end{figure}

We show representative examples of the various models in Figure~\ref{example}. Models 2-4 represent the most common kinds of artifacts we see in the data. Model 1 (the parabola) catches the case where a single strong artifact triggers a detection of a possible planet, and the other reported events show no strong signal. A box (Model 5) is the crudest possible model for the transit shape, but the second order details of ingress shape or limb darkening are not expected to be visible in the low SNR case of a single transit of a small planet.

The discontinuities in our models present a challenge for many optimization algorithms. We chose the Amoeba, or Nelder-Mead method \citep{Nelder65}, which does not require the first derivative of the model to be well behaved. We found the Amoeba often converged on a local minimum, so we repeated the fit for each model with a variety of initial conditions, and selected the best fitting result.

We occasionally found transits that lie very close to gaps in the data. The  fit is often poor when some of the expected data is missing, so we imposed a stringent requirement that no more than 25\% of the expected cadences in the selected snippet be missing or gapped before running the algorithm on that event.

We select the preferred model using the Bayesian Information Criterion \citep{Schwarz78}. This metric rewards models that fit the data with fewer parameters, and is given by

\begin{equation}
\mathrm{BIC} = -2\ln{\mathcal{L}} + k \ln{N}
\end{equation}

\noindent
where $k$ is the number of free parameters in the model and $N$ is the number of data points fit. $\mathcal{L}$ is likelihood of the fit, defined as

\begin{equation}
\mathcal{L} \propto \prod \exp \left[-\frac{1}{2}
\left( \frac{ y_i - f(t_i|\theta)}{\sigma} \right)^2
\right]
\end{equation}

\noindent
where $y_i$ is the value of the $i^{\rm th}$ data point, and $f(t_i|\theta)$ is the value of the model given a set of parameters $\theta$ and evaluated at the time of the $i^{\rm th}$ data point. $\sigma$ is the uncertainty assigned to each data point, calculated in a manner that is robust to outliers in the data

\begin{equation}
\sigma = 1.4826/\sqrt{2} \times \mad{y_{i+1} - y_i}
\end{equation}

The normalization ensures the computed value of $\sigma$ is consistent with the value expected if the values of the data points were drawn from a Gaussian distribution. The Gaussian assumption is frequently invalid for \kepler\ lightcurves, but any mismeasurement of $\sigma$ affects all model fits equally.

The model with the lowest value of the BIC is the preferred model. We define the fit score as the difference between the BIC value of the transit fit and the best fitting artifact model. Positive scores mean that the artifact model is preferred over the box model. \kepler\ catalogs adopt the principle of ``innocent until proven guilty'', erring on the side of including suspected false positives instead of incorrectly rejecting planet candidates. We therefore only reject events as false alarms if they have a score $\geq +10$. \citet{Kass95} argue that a difference in BIC value of 10 or more means the evidence against the disfavored model is very strong, {\add 
However, the choice of threshold is arbitrary; in the next section we describe how we tuned our algorithm on simulated transits to achieve the desired completeness (fraction of valid transits passed by the algorithm) at the chosen threshold, at the expense of high reliability (fraction of false alarms detected).}

 \begin{figure}
     \begin{center}
    \includegraphics[angle=0, scale=.37]{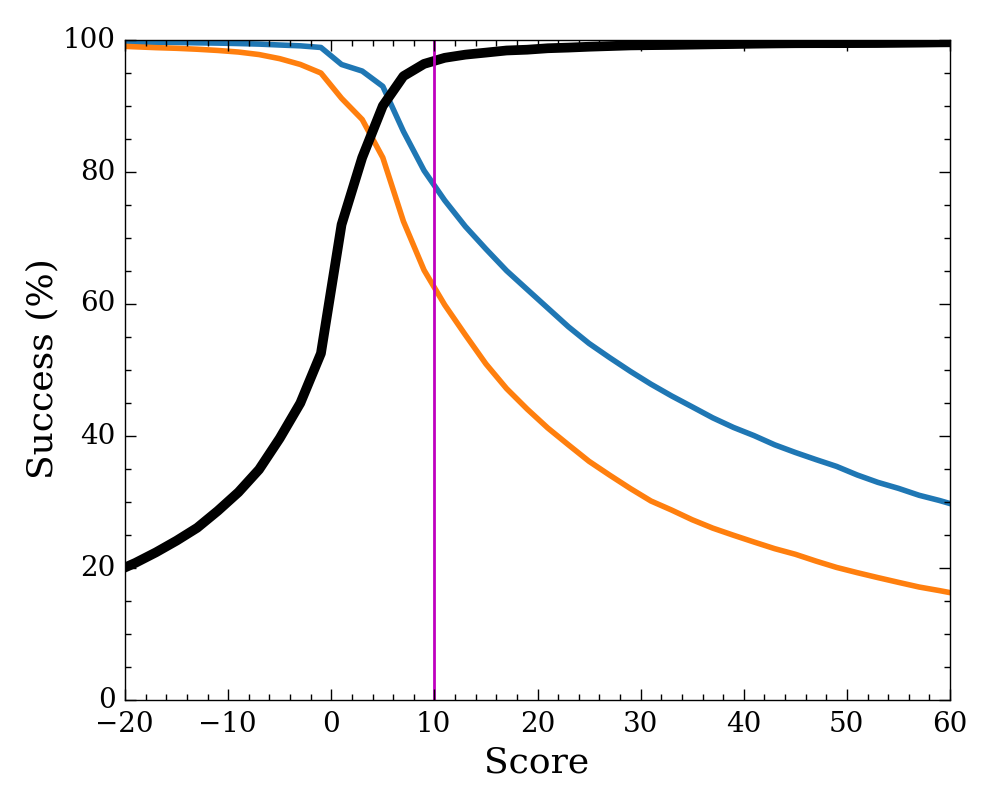}
    \caption{Performance of the algorithm on simulated data. The thick black line shows the percentage of simulated  individual transit events passed by Marshall as a function of the chosen threshold. The thin lines show the fraction of simulated artifacts failed at that same score. The more {\add positive} the score, the more the transit model is favored. We set a threshold of score $>$ +10 (vertical magenta line) to mark an event as a false alarm.
\label{sims}}
     \end{center}
 \end{figure}

\section{Performance }\label{performance}
{\add To evaluate the performance of our algorithm we ran it on the simulated transit set used to evaluate the completeness of the \kepler\ planet finding pipeline as described by \citet{Christiansen16} and based on the technique of \citet{Christiansen15}.} They injected a range of physically realistic transits for a range of planet sizes, orbital periods etc.
into the pixel level time-series data of all \kepler\ target stars using the method described in \cite{Christiansen15}.
This end-to-end modeling means that the effects on transit shape of the lightcurve generation \citep[PA,][]{Twicken10}, and systematic removal \citep[PDC,][]{Smith12} modules of the pipeline are properly accounted for in our simulation.

We selected for our study $10^{4}$ individual simulated transit events for planets with periods $>200$\,days and radii $<5$\,\Rearth\ that were recovered by the pipeline. We tuned our algorithm until $>95$\% of these individual events were passed. We then measured the performance of this tuned algorithm against simulated artifacts.

{\add As \citet{Christiansen16} did not inject artifact signals in their simulations, we perform our own. We add an artifact model to the lightcurves before co-trending, as produced by the PA module of the pipeline (the {\tt SAP\_FLUX} column of the MAST lightcurve file). We cotrend the simulated signal by fitting and removing the appropriate Covariant Basis Vectors produced by the PDC module of the pipeline \citep{Stumpe12}. These vectors represent the coarse systematic signals in \kepler\ lightcurves. They are available at MAST\footnote{\url{http://archive.stsci.edu/missions/kepler/cbv}}, are described in \S\,2.3.4 of the Archive Manual, and a tutorial on their use is given in \citet{Kinemuchi12}. PDC uses advanced techniques \citep{Smith12, Stumpe14} to apply the vectors to the data in an effort to prevent over-fitting of stellar variability. For our purposes, we capture the important behavior of PDC with the computationally simpler direct fits, so we apply those instead.
We run $10^4$ simulations over a range of targets and model parameters for each of the discontinuity and SPSD models. }


\begin{figure}
     \begin{center}
    \includegraphics[angle=0, scale=.37]{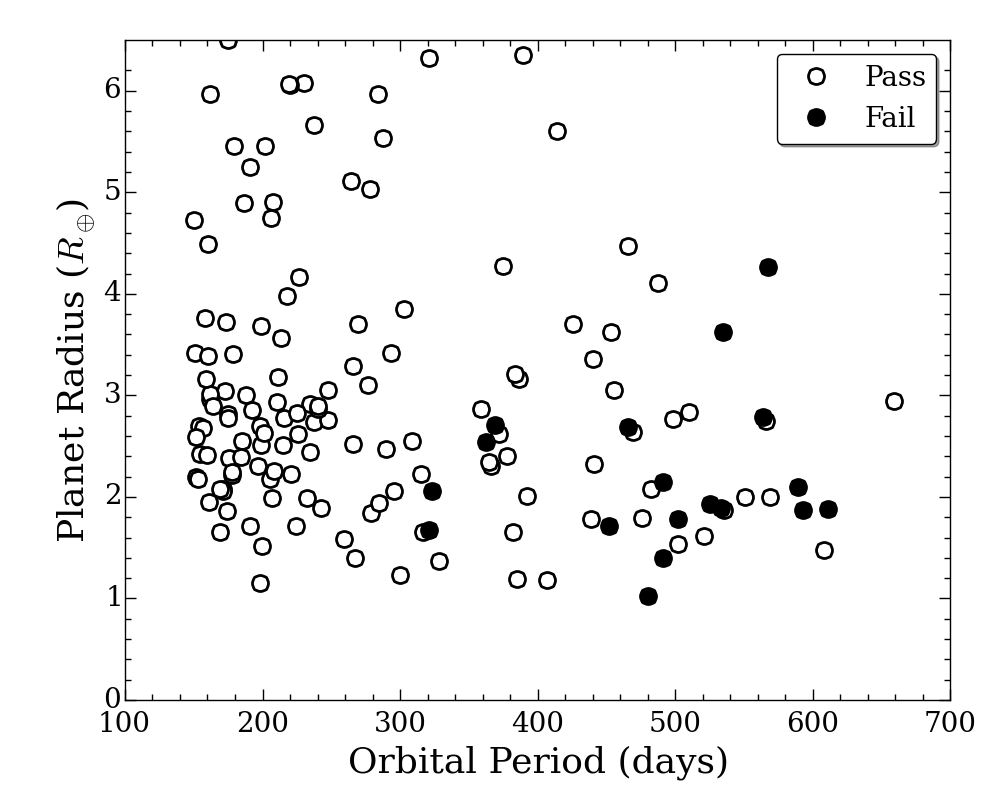}
    \caption{Long period KOIs that pass (open circles) and fail (filled circles) our test. KOIs that fail are clustered at long period and small radius, corresponding to low SNR detections. Almost all are detected with only three transits. Two KOIs detected at low SNR and large radii are not shown on this plot.
\label{q17Results}}
     \end{center}
 \end{figure}

We show our results in Figure~\ref{sims}. The thin blue and orange lines show our performance at identifying and rejecting systematic events in our simulations (discontinuities and SPSDs respectively) as a function of the threshold value. The vertical magenta line indicates the adopted threshold value of 10 (i.e the transit model has a BIC score at least 10 points higher than the most favored artifact model). At a threshold of 10 we reject 60-70\% of the injected events. The thick black line shows the number of injected transits that are {\em preserved} by the technique, which is $>96\%$. These results give us confidence that we can identify many of the false alarm events in the real data, while preserving the signal of most of the real planets. 

\section{Application to the \drtwofour\ Catalog}
The \drtwofour\ catalog of KOIs \citep{Coughlin15} incorporates results from Marshall. The KOI creation process is described in detail in \citet{Rowe15}. Briefly, a KOI number is assigned to a periodic signal in the lightcurve of a \kepler\ target that appears to be due to the transit or eclipse of an astrophysical body. KOI numbers are assigned based on a preliminary analysis and are sometimes assigned to other phenomena such as stellar variability or instrumental artifacts. Further vetting identifies some more of these artifacts as well as false positive signals due to eclipsing binaries. Any KOI is marked as a planet candidate unless there is conclusive evidence that it is not.

A KOI incorporates at least three events equally spaced in time, and Marshall deals with individual events. To apply Marshall to KOIs, we need to choose a disposition (planet candidate or false positive) based on the combined scores of individual events. We adopt the following rules:

\begin{enumerate}
\item Count $N_{\rm good}$, the number of transit events where $>75$\% of the expected cadences in the appropriate interval were collected, and where the computed score is $< +10$.

\item Count  $N_{\rm skip}$, the number of events where some data, but less than 75\% of the expected cadences, were collected. These events are not tested, because the gaps often severely bias the fits leading to inaccurate results. They frequently contain legitimate events so are assumed to pass by default.

\item If $N_{\rm good} + N_{\rm skip} \ge 3$ then the KOI passes, otherwise it fails. This rule is consistent with the mission requirement of needing at least three transits to claim the transit detection.

\end{enumerate}

\noindent

We apply this test to KOIs in the \drtwofour\ catalog with orbital periods $>$ 150\,days. We seed our fits with the transit parameters (period, epoch, duration etc.) from \citet{Seader15} and available at the NASA Exoplanet Archive \citep{Akeson13}. Although Marshall is tuned for low SNR events, and ignores second order effects on transit shape (such as ingress shape and limb darkening) we find it correctly identifies high SNR events as transits. {\add Our treatment of $N_{\rm skip}$ has a small effect on our results. From the set of \nKoisTested\ TCEs with periods $>$ 150\,days, and not already marked as false positives in the \drtwofour\ catalog, only 3 additional TCEs are marked as false positive if we require $N_{\rm good} \ge 3$.}

We show our results in Figure~\ref{q17Results} and Table~\ref{statusTable}. Of the \nKoisTested\ KOIs tested that were not otherwise failed by the Robovetter, 20 fail our test of having fewer than three valid transits (i.e $N_{\rm good} + N_{\rm skip} < 3$) mostly at periods $>$ 400\,days and radii $<$ 3\,\Rearth. Visual inspection confirms the artifact nature of all but two of these KOIs (K05805.01 and K02758.01). False positive identification in the \drtwofour\ catalog is made entirely by rule, so these KOIs are marked as false positive in the final catalog in \citet{Coughlin15} even though visual inspection identifies them as legitimate candidates.

\section{Discussion}

One limitation of our technique is that we must make an assumption as to how the lightcurve would look in the absence of a transit at a given epoch. We choose a simple model of a parabola fit to a small portion of the out-of-transit lightcurve that we find works well in practice, but there are cases of stars that exhibit such rapid variability  that we have difficulty measuring an accurate continuum. This causes some of our events to be misidentified. A more sophisticated estimate of the continuum would likely improve performance, but would have to account for impulsive (i.e. short duration) spacecraft events, as well as the variability of the star itself.

\begin{figure}
     \begin{center}
    \includegraphics[angle=0, scale=.37]{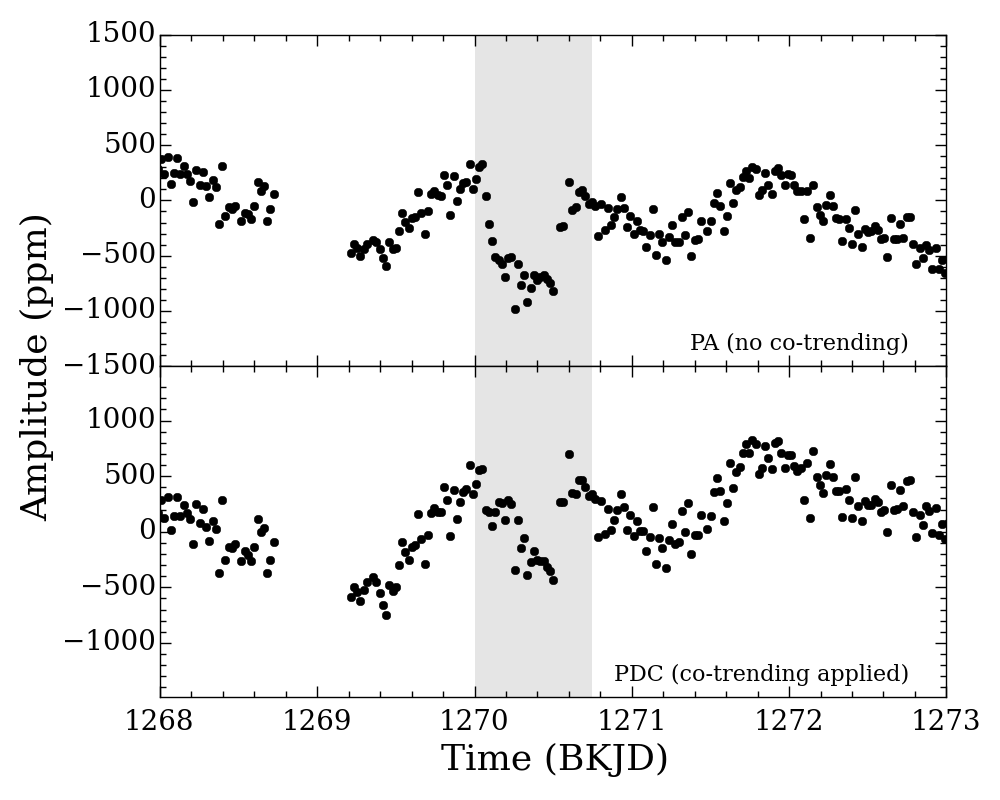}
    \caption{An example of how PDC can deform the shape of a transit. The
top panel shows the lightcurve of Kepler Id 6520870 with a
simulated transit (highlighted by the gray bar). The lower
panel shows this same data after co-trending by PDC against data
from other nearby stars to remove coarse trends. 
\label{pdcMangle}}
     \end{center}
 \end{figure}

For the 4\% of injected transits that were rejected, we inspected the data to understand why they failed. We show an example in Figure~\ref{pdcMangle}. The top panel shows the injected event, while the bottom panel shows the lightcurve after long term trends were removed by PDC. In this case, the shape of the transit was deformed to make it look more like a systematic effect. For stars without rapid variations in the lightcurve, this is the most common reason why simulated planet candidates were misidentified.
However, attempts to use the lightcurves without the detrending from PDC had significantly worse performance due to the more complicated structure in the out-of-transit flux.

Although the artifact types we test for create most of the artifacts we are aware of for small, long period planet candidates, there are presumably other sources of systematics not yet accounted for that decrease the sample of planet candidates still further. In addition, our simulations suggest we only find two thirds of our injected false alarms This suggests as many as \nFaMissed\ more false alarms of the kind we tested for remain in the \drtwofour\ catalog.
{\add There are 37 planet candidates with periods longer than 300\,days in the catalog.
If, similar to the identified false alarms, the uncaught systematics also have periods $>300$\,days, then we expect no more than 27 (or 73\%) of these long period candidates are actually transits. This places a rough upper bound on the reliability of the catalog for small, long period planets.}


{\add The \kepler\ Robovetter emphasizes completeness over reliability. Ensuring that as many planets as possible are included as candidates in the catalog is a higher priority than removing as many false positives as possible. Users of the catalog who place a higher value on reliability may prefer to set their rejection threshold at at lower BIC score, at the expense of a decrease in the completeness of their sample. To this end we provide the Marshall score for each KOI tested in Table~\ref{statusTable}. 

Finally, we note that the score given in Table 1 is not the probability a KOI is due to a transit. Even KOIs with strong BIC scores may be due to non-planetary signals, such the highly diluted signal of an eclipsing binary system with small angular separation from the target.
}


\section{Conclusion}
We present a new technique to identify systematic signals in \kepler\ data masquerading as planet candidates. The algorithm looks at each individual transit event and decides if the shape of that event is more likely a transit or one of a few known artifacts. We apply the algorithm to the DR24 planet catalog of \citet{Coughlin15} and find it rejects \nKoisFail\ small, long period KOIs otherwise marked as planet candidates by the Robovetter.

\acknowledgements
Funding for this Discovery mission is provided by NASA’s
Science Mission Directorate.
All of the data presented in this paper were obtained from the Mikulski Archive for Space Telescopes (MAST). STScI is operated by the Association of Universities for Research in Astronomy, Inc., under NASA contract NAS5-26555. Support for MAST for non-HST data is provided by the NASA Office of Space Science via grant NNX13AC07G and by other grants and contracts.
This research has made use of the NASA Exoplanet Archive, which is operated by the California Institute of Technology, under contract with the National Aeronautics and Space Administration under the Exoplanet Exploration Program.

{\it Facilities}: \kepler

\bibliographystyle{apj}
\bibliography{q1q16, marshall}

\input{table1_short}

\end{document}

%% file: table1_short.tex
\begin{deluxetable}{lrrrccr}
\tablewidth{0pt}
\tabletypesize{\normalsize}
\tablecaption{KOIs examined by Marshall}

\tablehead{
    \colhead{KOI}&
    \colhead{Kepler Id}&
    \colhead{TCE}&
    \colhead{NTrans.} &
    \colhead{Period (days)} &
    \colhead{Radius (R$_{\oplus}$)} &
    \colhead{Score} \\
}
\startdata

K00998.01 & 1432214 & 1 & 9 & 161.788000(36) & 30.140(49) & $-2.1 \times 10^{4}$\\
K01099.01 & 2853093 & 1 & 9 & 161.52800(34) & 5.97(76) & $-5.5 \times 10^{1}$\\
K01788.02 & 2975770 & 2 & 4 & 369.0790(26) &  2.7(2.4) & $+2.7 \times 10^{1}$\\
K03549.01 & 2307206 & 1 & 7 & 204.029000 0 & 44.140000 0 & $-1.4 \times 10^{5}$\\
K03674.01 & 2446623 & 1 & 8 & 175.066000(66) &  28.6(7.1) & $-4.2 \times 10^{3}$\\
K03681.01 & 2581316 & 1 & 7 & 217.832000(77) & 11.410(19) & $-8.0 \times 10^{4}$\\
K03709.01 & 2576107 & 1 & 7 & 205.58300(16) &  21.0(2.6) & $-2.8 \times 10^{3}$\\
K04959.01 & 2987433 & 1 & 4 & 420.92900(10) &  49.4(1.7) & $-5.0 \times 10^{5}$\\
K06254.01 & 1433531 & 1 & 3 & 567.676(13) & 4.27(46) & $+3.1 \times 10^{1}$\\
K06295.01 & 2856960 & 1 & 7 & 204.30400(66) & 13.730(62) & $-1.4 \times 10^{3}$\\

\enddata
\tablecomments{\label{statusTable}
Summary of planet candidate KOIs from \drtwofour\ with periods greater than 150\,days examined by Marshall.
 Kepler Id is the unique identifier of the target star in the \kepler\ input catalog. TCE indicates the order in which this KOI was found around the target star by the \kepler\ pipeline in Q1-Q17. NTrans. refers to the number of observed transits as measured by the pipeline. The uncertainty in the last two digits of period and radius are given in parentheses. Period and radius values are taken from \citet{Seader15} and may differ slightly from the final values in \citet{Coughlin15}.
Score represents the confidence Marshall assigns to a KOI. Negative scores indicate high confidence that a KOI is due to a transit while positive scores indicate lower confidence. KOIs with scores $>$10 are deemed to be false positives. The score given in the table is equal to the score of the 3$^{\mathrm{rd}}$ strongest individual event for the KOI, which indicates how close the KOI is to being marked as a false positive. 
(This table in available in its entirety in the on-line supplementary materials. A portion is shown here for guidance regarding its form and content.)}
\end{deluxetable}